\documentclass[global,twocolumn]{svjourArXiv}

\usepackage{graphicx}

\newcommand{\etal}{\mbox{et~al.}}

\newcommand{\mum}{\ensuremath{\mu\mathrm{m}}}
\newcommand{\mm}{\ensuremath{\mathrm{mm}}}
\newcommand{\Done}{\ensuremath{\mathrm{D}_{1}}}

\newcommand{\rf}{\ensuremath{\mathrm{rf}}}

\newcommand{\Hz}{\ensuremath{\mathrm{Hz}}}
\newcommand{\uT}{\ensuremath{\mu\mathrm{T}}}
\newcommand{\nT}{\ensuremath{\mathrm{nT}}}
\newcommand{\pT}{\ensuremath{\mathrm{pT}}}
\newcommand{\fT}{\ensuremath{\mathrm{fT}}}
\newcommand{\fTHz}{\ensuremath{\mathrm{\fT}/\sqrt{\Hz}}}
\newcommand{\NEM}{\ensuremath{\mathrm{NEM}}}

\newcommand{\AnaFF}{\ensuremath{A_{FF'}}}
\newcommand{\AnaFFus}{\ensuremath{A_{43}}}
\newcommand{\Fpol}{\ensuremath{\braket{F_z}}}
\newcommand{\Pdet}{\ensuremath{P_{det}}}

\newcommand{\PAC}{\ensuremath{P^{LIA}_{det}}}
\newcommand{\Pout}{\ensuremath{P_{L}}}
\newcommand{\Pin}{\ensuremath{P_{in}}}

\newcommand{\Nin}{\ensuremath{N_{in}}}

\newcommand{\NAC}{\ensuremath{N^{LIA}_{det}}}
\newcommand{\NDC}{\ensuremath{N^{DC}_{det}}}

\newcommand{\Qe}{\ensuremath{{Q}_{\kern-0.2em E}}}

\def\ket#1{\mathinner{|{#1}\rangle}}
\def\braket#1{\mathinner{\langle{#1}\rangle}}

{\catcode`\|=\active
  \gdef\Braket#1{\left<\mathcode`\|"8000\let|\bravert
  {#1}\right>}}
\def\bravert{\egroup\,\vrule\,\bgroup}

\newcommand{\sGamma}[1]{\ensuremath{\Gamma_{\kern-0.23em #1}}}

\journalname{Applied Physics B}
\begin{document}

\title{A large sample study of spin relaxation and magnetometric
  sensitivity of paraffin-coated Cs vapor cells}

\author{N.~Castagna
\thanks{\emph{Email address:} natascia.castagna@unifr.ch}
\and G.~Bison
\thanks{\emph{Present address:} Universit\"atsklinikum Jena, Jena, Germany.}
\and G.~Di~Domenico
\thanks{\emph{Present address:} Laboratoire Temps-Fr\'equence, Universit\'e de Neuch\^atel, Neuch\^atel, Switzerland.}
\and A.~Hofer
\and P.~Knowles
\and C.~Macchione
\and H.~Saudan
\and and A.~Weis}

\institute{Universit\'e de Fribourg, Chemin du Muse\'e 3,
CH--1700, Fribourg, Switzerland}
\date{Received: date / Revised version: date}

\maketitle

\begin{abstract}

We have manufactured more than 250 nominally identical paraffin-coated
Cs vapor cells (30~mm diameter bulbs) for multi-channel atomic
magnetometer applications.  We describe our dedicated cell
characterization apparatus.  For each cell we have determined the
intrinsic longitudinal, $\sGamma{01}$, and transverse, $\sGamma{02}$,
relaxation rates.  Our best cell shows
$\sGamma{01}/2\pi\approx~0.5~\Hz$, and
$\sGamma{02}/2\pi\approx~2~\Hz$.  We find a strong correlation of both
relaxation rates which we explain in terms of reservoir and spin
exchange relaxation.  For each cell we have determined the optimal
combination of \rf{} and laser powers which yield the highest
sensitivity to magnetic field changes.  Out of all produced cells,
90\% are found to have magnetometric sensitivities in the range of
9~to 30~\fTHz.  Noise analysis shows that the magnetometers operated
with such cells have a sensitivity close to the fundamental photon
shot noise limit.

\end{abstract}

\section{Introduction}

Spin polarized alkali vapors prepared by optical pumping have been
used since 50 years for fundamental studies in atomic physics and
applications thereof \cite{Budker:2002:RNM}.
The achieved sensitivities depend mainly on the (transverse)
lifetime $T_{2}$ of the spin coherence in the vapor, and, to a
lesser extend, on the (longitudinal) lifetime $T_{1}$ of the spin
polarization.
Those lifetimes are related to corresponding relaxation rates by
$\sGamma{i}=T_{i}^{-1}$.
To assure a long-lived spin polarization, the vapor cells are either
filled with a buffer gas mixture or are left evacuated while applying
an anti-relaxation coating on the walls.
In the first case, the buffer gas in the cell confines the atoms to a
diffusion-limited volume and thus reduces the rate of depolarizing
wall collisions.
In the second case, a thin film of paraffin or similar substance
applied to the cell wall reduces the collisional sticking time with
the wall and thereby the dephasing interactions with magnetic
impurities embedded in the walls.

Alkali vapors in paraffin-coated cells were introduced in 1958
\cite{Robinson:1958:PSS} and have since been widely applied in atomic
physics spanning applications from magnetometers
\cite{DiDomenico:2007:SDR,Weis:2005:LBP,Budker:2007:OM},
over slow light studies
\cite{Klein:2006:SLP}, to spin-squeezing
\cite{Fernholz:2008:SSA}, and light-induced atomic desorption (LIAD)
\cite{Alexandrov:2002:LID,Gozzini:2008:LIS} studies.

Our group develops atomic magnetometers for the accurate measurement
of small changes in already weak fields (typically 10\% of the earth's
field \cite{Bison:2005:OPO}), a technique that we currently apply to
the measurement of the faint magnetic fields produced by the beating
human heart) \cite{Bison:2003:LPM,Weis:2006:TDR,Hofer:2008:HSO} and
for magnetic field measurement and control in the search for a neutron
electric dipole moment \cite{Groeger:2006:HSL,Ban:2006:TNM}.
Both experiments call for a large number (50 to 100) of individual
sensors to be operated simultaneously.
Although buffer gas cells were used in our initial work
\cite{Bison:2003:LPM}, we currently focus on paraffin-coated cells
that have a reduced sensitivity to magnetic field gradients because of
motional narrowing and to temperature effects compared to buffer gas
cells \cite{Andalkar:2002:HRM,Vanier:1989:QPA}.

In order to fulfill the requirements of the mentioned experiments we
have initiated a large scale production of cells that has yielded over
250 cells in the past year.  We have developed an automatic cell
characterization facility for determining the quality and
reproducibility of the cell coatings.  In this work we describe this
characterization facility in detail and report results (intrinsic
relaxation times, intrinsic magnetometric sensitivity) based on
significant cell statistics.  A comparative study of a small sample of
paraffin-coated cells produced over four decades was reported in
\cite{Budker:2005:MTN}.  To our knowledge our present study involves
the largest sample of coated cells ever compared.

\section{Cell production}
\label{sec:cells}

\begin{figure}[t]
\centerline{\includegraphics*[width=\linewidth]{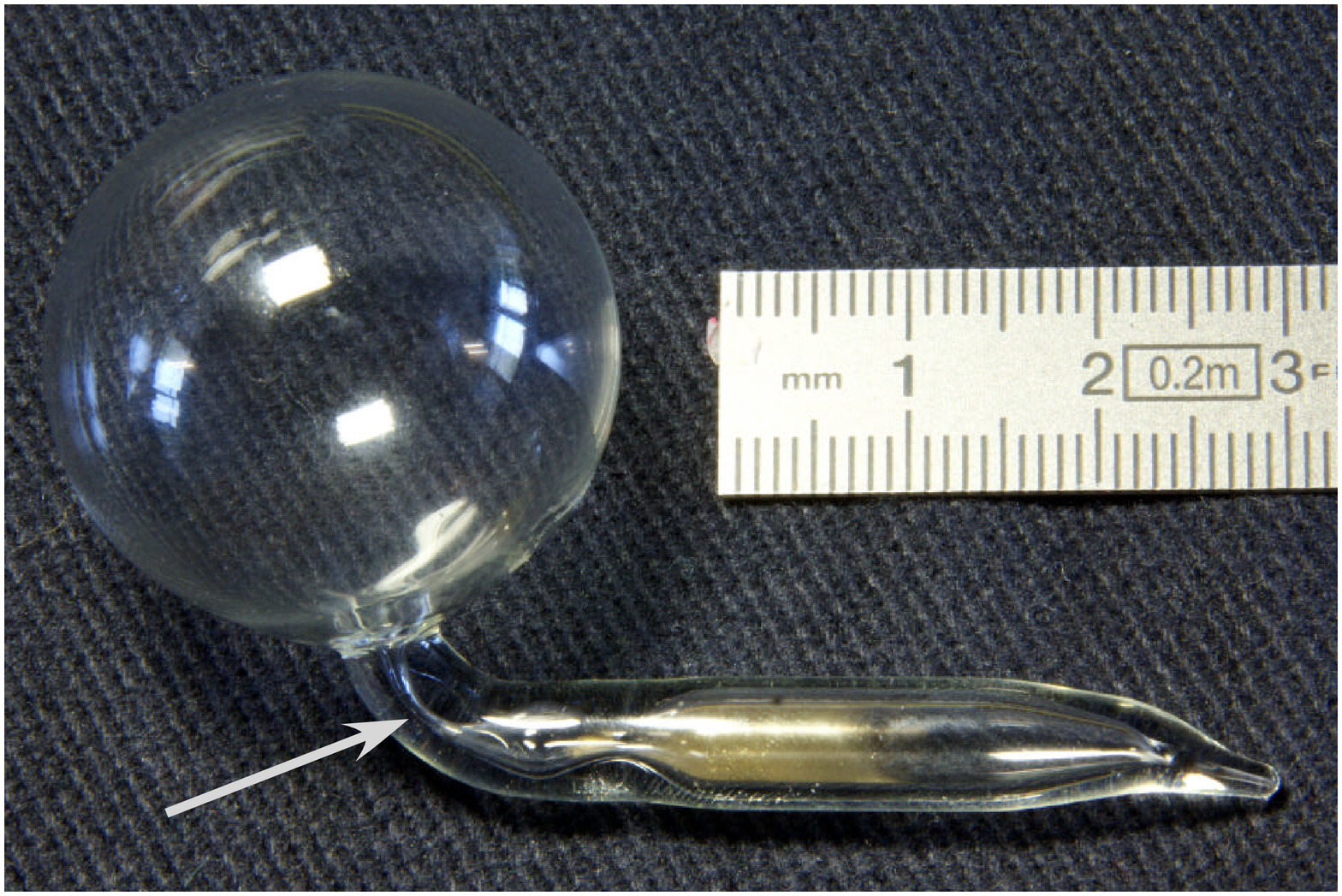}}
\caption{(Color online) Paraffin-coated Cs vapor cell. The small
  amount of the solid alkali metal is well visible in the sidearm.
  The arrow points to the capillary which reduces depolarizing
  collisions of vapor atoms from the cell with the solid Cs.}
\label{fig:cell}
\end{figure}

The paraffin-coated glass cells are manufactured in our institute.
Pyrex is formed into a spherical bulb (inner diameter of $\approx
28~\mm$, wall thickness of 1~\mm) that is connected to a sidearm
consisting of a Pyrex tube with 4~\mm{} inner (7~\mm{} outer)
diameter, which acts as a reservoir to hold the droplet of solid
cesium after coating, filling, and sealing the cell
(Fig.~\ref{fig:cell}).  The metallic Cs is the source for the
saturated Cs vapor filling the cell.  Near the cell proper, the
sidearm is constricted into a capillary with a design diameter of
0.75(25)~\mm{} that reduces spin depolarizing collisions with the bulk
Cs in the sidearm.

A typical coating and filling process takes about one week.  Ten cells
are mounted on a glass structure together with a paraffin containing
reservoir and a Cs metal containing ampule, both isolated from the
vacuum system by break-seals.
The system is connected to a turbomolecular pump stand via a liquid
nitrogen cold trap and all coating and filling steps are performed in
a vacuum below $10^{-7}$~mbar.
Prior to coating, the whole structure is baked for 5~hours at
$370{}\,^\circ$C.

The coating process is similar to the one reported in
\cite{Alexandrov:2002:LID,Gozzini:2008:LIS}.
Our current choice of coating material is a commercial paraffin,
Paraflint $H_{1}$, from Sasol Wax American Inc.
After baking the system, the break-seal of the paraffin reservoir
is broken by a piece of iron sealed in a glass bead (``hammer'')
manipulated from the outside by a permanent magnet.
The wax is deposited onto the cell walls by heating the paraffin
reservoir.
During the coating procedure the pressure rises to $9\times
10^{-7}$~mbar, and the cell is kept isolated from the cesium
containing ampule.
Once the cell is coated, the same hammer is used to break the seal of
the Cs ampule and a thin film of metallic Cs is distilled into the
cell's sidearm by heating the Cs ampule, after which the end of the
sidearm is sealed off.
During Cs distillation the pressure rises to $3\times10^{-7}$~mbar,
and at the end of filling the cells are pumped down to a pressure
below $10^{-7}$ mbar before being sealed.
The filled cells are activated by heating them in a oven at
$80{}\,^\circ$C for 10~hours, while assuring that the sidearm is kept
at a sightly lower temperature.
In this way we produce 10 coated cells in one week.

\section{The cell characterization setup}
\label{sec:experiment}

\begin{figure}[b]
\centerline{\includegraphics*[angle=0,width=\linewidth]{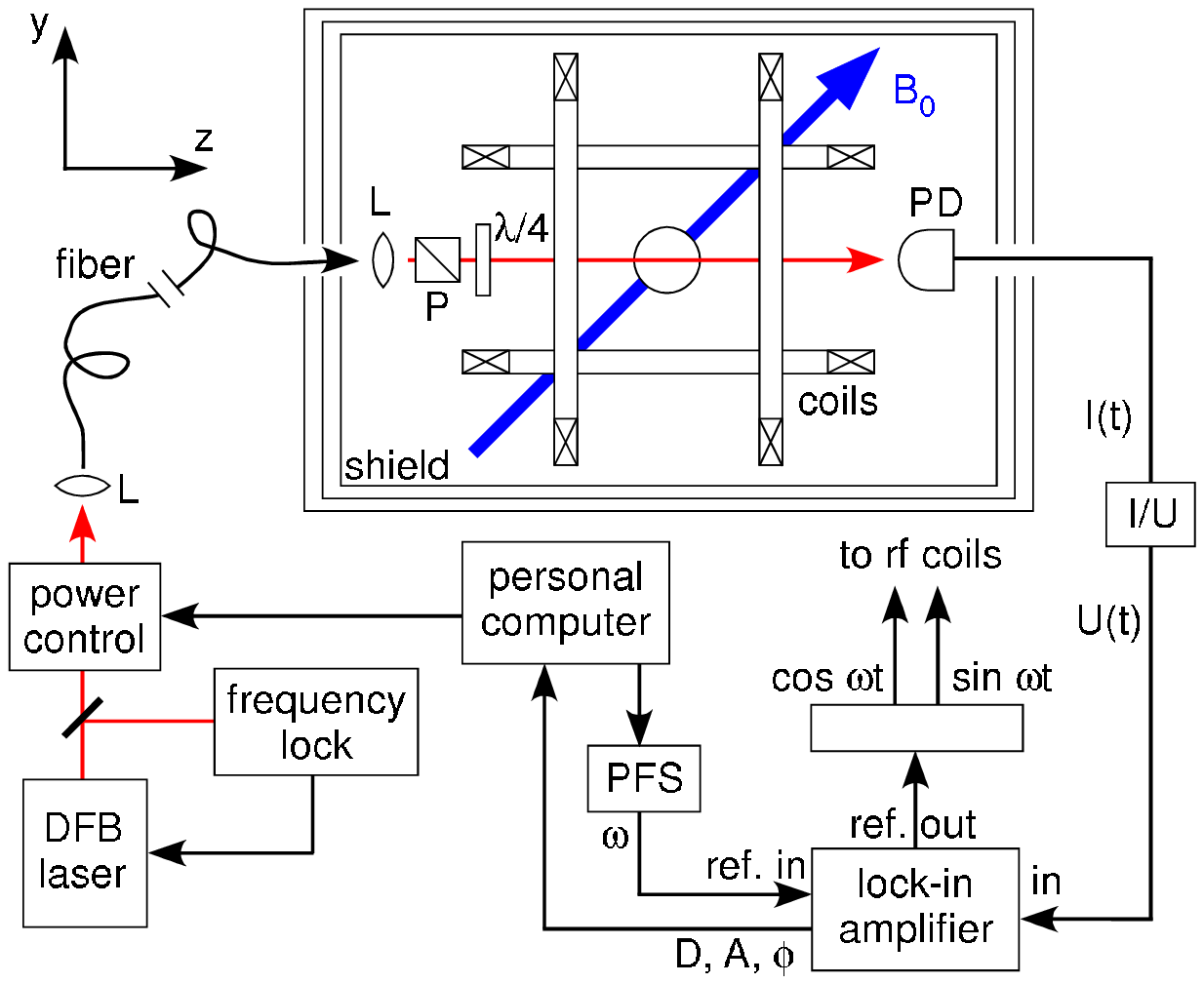}}
\caption{The cell testing apparatus.  Frequency stabilized laser light
  is carried by a multimode fiber into a threefold magnetic shield (L:
  lenses). Circular polarization is created by a polarizer (P) and a
  quarter-wave plate ($\lambda/4$).  The transmitted power is recorded
  by a photodiode (PD) and the modulated light power components are
  extracted by a lock-in amplifier.  A personal computer controls the
  light power, performs scans of the frequency $\omega$ via a
  programmable frequency synthesizer (PFS, Stanford Research model
  SR345), and records the lock-in signals.}
\label{fig:expsetup}
\end{figure}

Following manufacture, each cell undergoes a characterization
procedure in a dedicated experimental apparatus for determining the
relevant parameters that indicate its magnetometric properties.
Our current magnetometers use the technique of optically detected
magnetic resonance in the Double Resonance Orientation
Magnetometer or DROM  configuration (notation introduced in
\cite{Weis:2006:TDR}), also called M$_x$-configuration
\cite{Aleksandrov:1995:LPS,Groeger:2006:HSL}. The underlying
theory will be addressed below.
It was thus a natural choice to use the same technique for the
dedicated cell testing facility.

\subsection{Experimental setup and signal recording}

The experimental apparatus is shown in Fig.~\ref{fig:expsetup}.
The laser source is a DFB laser ($\lambda=894$~nm) whose frequency is
actively stabilized to the $4 \Rightarrow 3$ hyperfine component of
the Cs \Done{} transition using the dichroic atomic vapor laser lock (DAVLL)
technique \cite{Corwin:1998:FSD}.
The light is carried by a 400~\mum{} diameter multimode fiber into a
three-layer mu-metal magnetic shield that contains the actual double
resonance setup.
Prior to entering the fiber, the laser power, $P_L$, is
computer-controlled via a stepper-motor driving a half-wave plate
located before a linear polarizer.
The light leaving the fiber is collimated and passes a linear
polarizer followed by a quarter-wave plate to create circular
polarization before entering the Cs cell.
The fiber is wound into several loops so that the exiting light is
completely depolarized, thus avoiding vibration related polarization
fluctuations that translate into power fluctuations after the
polarizer.

The paraffin-coated Cs cell to be characterized is placed in the
center of the magnetic shields where three pairs of Helmholtz coils
and three pairs of anti-Helmholtz coils compensate residual stray
magnetic fields and gradients, respectively.
A static magnetic field $B_0$ with an amplitude of a few \uT{} is
applied in the $yz$-plane at $45^{\circ}$ with respect to the laser
beam direction, $\hat{k}=\hat{z}$.
The transmitted light power is recorded by a nonmagnetic photodiode
and then amplified.
Absorbed laser light pumps the Cs atoms into the nonabsorbing (dark)
$\ket{F{=}4, M_F{=}3,4}$ magnetic sublevels, thereby creating a
vector spin polarization (orientation) $P_z\propto \braket{F_z}$.
A small magnetic field \rf-field $B_{1}(t)$ of a few~\nT, constant
in amplitude, but rotating at frequency $\omega$, is applied in
the plane perpendicular to $B_0$.
The choice of a rotating, rather than a linearly polarized,
oscillating field is used to suppress magnetic resonance transitions
in the $F{=}3$ state \cite{DiDomenico:2006:ESL}.
$B_{1}(t)$ drives magnetic resonance transitions between adjacent
sublevels in the $F{=}4$ hyperfine state, whose Zeeman degeneracy is
lifted by the static magnetic field $B_0$.
For a properly oriented magnetic field $\vec{B}_0$ the transmitted
light power will be modulated at the rotation frequency $\omega$.

When $\omega$ is close to the Larmor frequency $\omega_L = \gamma_F
B_0$, where $\gamma_F \simeq 2\pi \cdot 3.5~\Hz/\nT$ is the Cs ground
state gyromagnetic factor, a resonance occurs in the absorption
process, manifesting itself in both the amplitude and phase of the
light power modulation.
The corresponding in-phase, quadrature, and phase signals are
extracted by means of a lock-in amplifier (LIA, Stanford Instruments,
model SR830) whose output signals are read by a personal computer.
The rotating field frequency is generated by a computer controlled
programmable synthesizer.
The computer varies this frequency, $\omega$, by a linear ramp in the
range of $\pm~2\pi\cdot 100~\Hz$ around the Larmor frequency during a
scan time of 40~s.
A~dedicated electronics box generates from this AC voltage two
$90^\circ$ dephased AC currents that drive two perpendicular coil
pairs (not shown in Fig.~\ref{fig:expsetup}) producing the rotating
field $B_1(t)$. \\
\indent The characterization of each individual cell consists in the recording
of resonance spectra for a set of 12 selected (and computer
controlled) values of the laser power $P_L$ in the range of 1~to
$12~\mu$W.
It is difficult to determine the absolute laser intensity for a given
laser power $P_L$, because of the (asymmetric) transverse beam
profiles and their modification by the cell's spherical shape.
We therefore quantify the light intensity in terms of the laser power
$P_L$, to which it is proportional.
Note that $P_L$ used below refers to the power measured after the cell with
the laser frequency resonant with the $4{\rightarrow}3$ Cs~\Done{}
transition and the \rf{} power off.
A typical automated characterization run, including insertion of the
cell into the apparatus, takes 10 minutes.
Data analysis is performed by a semi-automatic dedicated
Mathematica\cite{Mathematica52} code, which takes another 5~minutes.
In a regular working day it is thus possible to characterize 30 to 40
cells.

\subsection{DROM theory}
\label{sec:theory}

A modulation of the transmitted power only occurs when the static
magnetic field $B_0$ is neither parallel nor perpendicular to the
direction of light propagation.
In that case the transmitted light power has components that oscillate
in phase, $D_{\omega}$, and in quadrature, $A_{\omega}$, with respect
to the rotating field
\begin{equation}
B_1(t)= \frac{\Omega_{\rf}}{\gamma_F}\, e^{i\,\omega t}\,.
\label{eq:in-phase}
\end{equation}

The in-phase and the quadrature components depend on the detuning,
$\delta=\omega-\omega_0$, between the driving, $\omega$, and the
Larmor, $\omega_0$, frequencies.
The dependence of $D_{\omega}$ and $A_{\omega}$ on $\delta$ are
dispersive and absorptive Lorentzians given by \cite{Bison:2005:OPO}
\begin{eqnarray}
D_\omega(\delta) = -\eta \Fpol \sin{(2\theta)}
                      \frac {\Omega_{\rf} \delta}
   { \delta^2+ \sGamma{2}^{2}+\frac{\sGamma{2}}{\sGamma{1}} \Omega^2_{\rf} }
\nonumber \\
A_\omega(\delta) = -\eta \Fpol \sin{(2\theta)}
                      \frac {\Omega_{\rf} \sGamma{2}}
   { \delta^2+ \sGamma{2}^{2}+\frac{\sGamma{2}}{\sGamma{1}} \Omega^2_{\rf} }
\label{eq:signals}
\end{eqnarray}
where $A_0=\eta \Fpol \sin{(2\theta)}$ is a common signal amplitude
that depends --- via the spin polarization \Fpol{} created by optical
pumping and the detection of the polarization's precession via light
absorption --- on the laser power $P_L$.
The calibration constant $\eta$ includes all of the apparatus
constants such that $D_\omega(\delta)$ and $A_\omega(\delta)$ are
measured in Volts.
With respect to Fig.~\ref{fig:expsetup}, $U(t) = D_\omega(\delta)
\cos{\omega t} + A_\omega(\delta) \sin{\omega t}$.
The phase $\phi_{\omega} (\delta)$ between the drive and the power
modulation
\begin{equation}
\phi_{\omega} (\delta) = +\arctan{\left(\frac{\sGamma{2}}{\delta}\right)}\,,
\label{eq:phase}
\end{equation}
depends also on the detuning $\delta$.
The expressions (\ref{eq:signals}) and (\ref{eq:phase}) are valid
for atomic media with an arbitrary ground state angular momentum,
as may be shown easily by a theoretical treatment analogous to the
discussion of the signals in the DRAM (double resonance alignment
magnetometer) geometry presented in \cite{Weis:2006:TDR}.
In the above expressions, $\theta$ is the angle between the
applied magnetic field $B_0$ and the laser beam propagation
direction $\vec{k}$.

\begin{figure}[t]
\centerline{\includegraphics[width=\columnwidth]{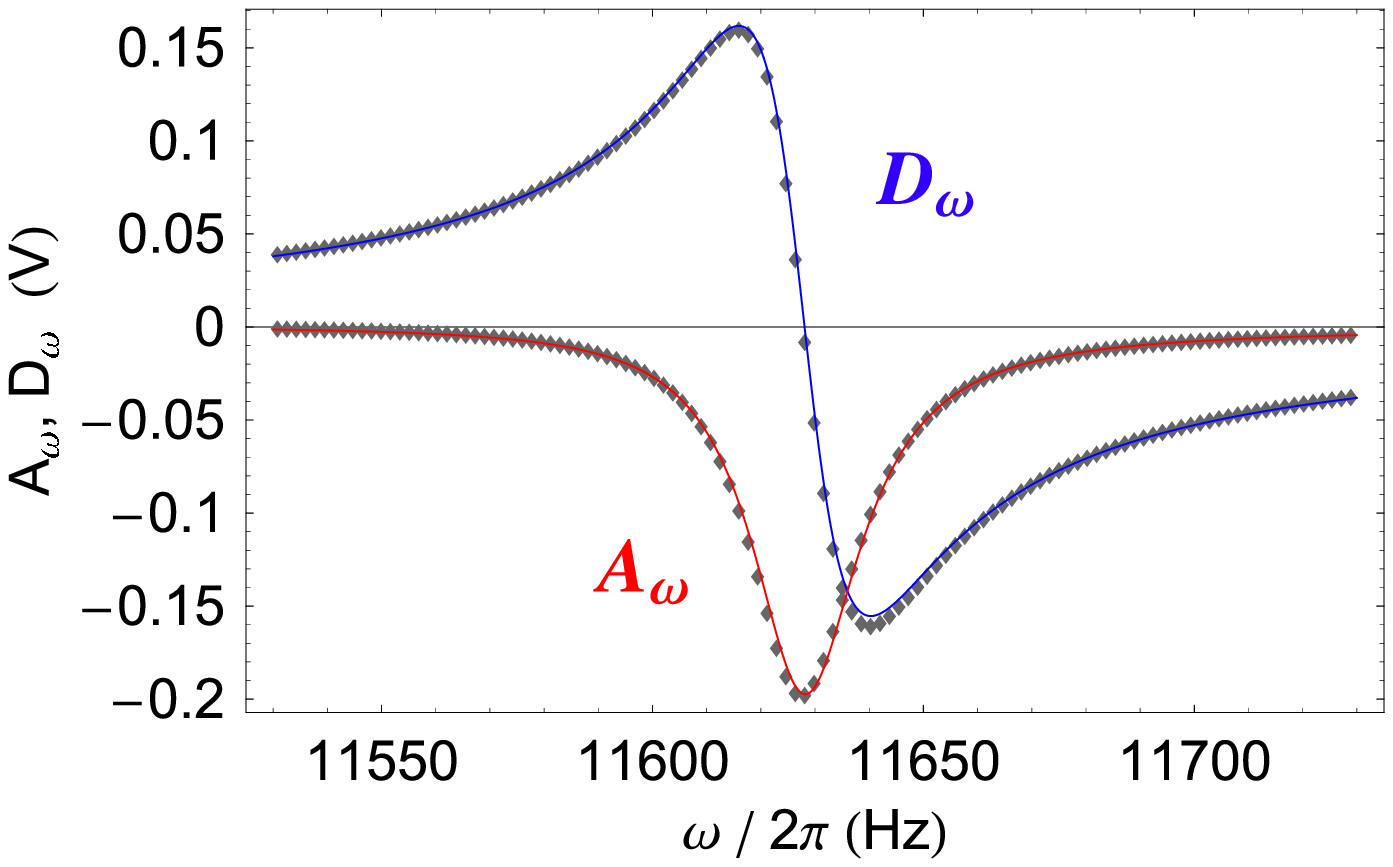}}
\centerline{\includegraphics[width=\columnwidth]{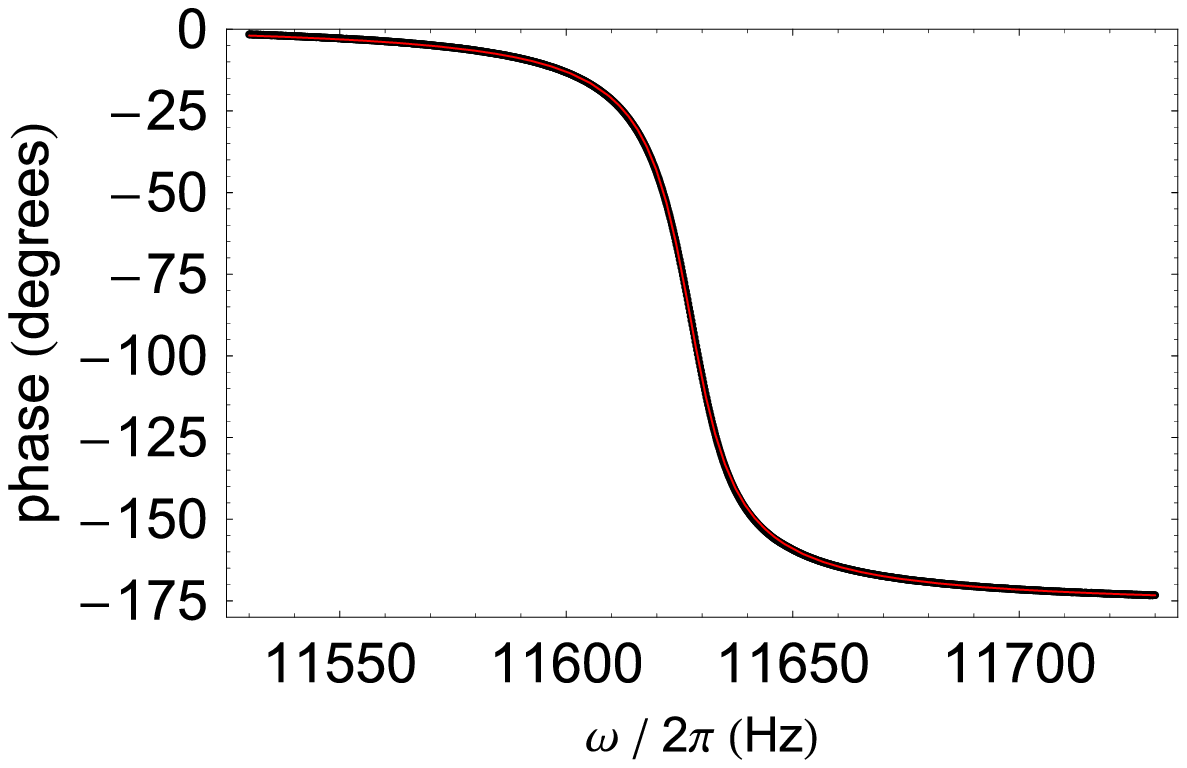}}
\caption{(color online) Lock-in demodulated magnetic resonance
  signals. Top: The dispersive signal (blue) represents the in-phase
  component $D(\omega)$ and the absorptive signal (red) the quadrature
  component $A(\omega)$.  Bottom: Phase signal $\phi(\omega)$.
  Experimental points are shown together with lines fitted according
  to (\protect\ref{eq:signals})--(\protect\ref{eq:phase}).  All signals
  were recorded at $B_0 \simeq 4~\uT$ ($\omega_0 \simeq 2\pi \cdot
  11640~\Hz$), with $P_L=6~\mu$W, and $B_{1}=1.3~\nT$.}
\label{fig:signals}
\end{figure}

\subsection{Signal analysis}
\label{sec:recording}

Since the resonance signals are extracted by a lock-in amplifier, and
since it is experimentally difficult to precisely determine the phase
of the rotating field (and hence the phase difference between that
field and the modulation of the photocurrent), the signals produced by
the lock-in amplifier are superpositions of the absorptive and
dispersive lineshapes $A_{\omega}(\delta)$ and $D_{\omega}(\delta)$.
Using the fitting procedure described in detail in
\cite{Bison:2005:OPO} it is possible to extract the pure absorptive
and dispersive components.
For fitting the theoretical lineshapes the combined apparatus
constants $A_0\equiv\eta \Fpol \sin{(2\theta)}$ is taken as one fitting
parameter, with $A_0$ measured in Volts.
Other parameters are the relaxation rates $\sGamma{1}$ and
$\sGamma{2}$, the resonance frequency $\omega_0$, an unknown overall
phase, as well as weighting factors of the absorptive and dispersive
components.
The Rabi frequency $\Omega_{\rf}$ can be easily calibrated as
described in \cite{DiDomenico:2007:SDR} and a fixed numerical value is
used when fitting (\ref{eq:signals}) and (\ref{eq:phase}).

Typical resonance lineshapes of the in-phase, quadrature, and phase
signals are shown in Fig.~\ref{fig:signals}, together with the fitted
theoretical shapes (\ref{eq:signals}) and (\ref{eq:phase}).
Fitting the absorptive and dispersive spectra by (\ref{eq:signals})
with the relaxation rates $\sGamma{1}$ and $\sGamma{2}$ as free
parameters yields a strong correlation between the two rates in the
$\chi^2$-minimizing algorithm, with corresponding large uncertainties
in the numerical values.
We have therefore opted for the following fitting procedure.
In a first step, we use the fact that the phase does not depend on
$\sGamma{1}$ and fit the dependence $\phi({\omega})$ given by
(\ref{eq:phase}) to the data.
The resulting $\sGamma{2}$ value is then used as a fixed parameter in
the subsequent simultaneous fit of the absorptive and dispersive
lineshapes to infer $\sGamma{1}$.
In this way we obtain $(\sGamma{1}, \sGamma{2})$-pairs for each value
of the laser power $P_L$.
In addition, the fits yield the overall signal amplitude $A_0$.

\begin{figure}[t]
\centerline{\includegraphics[angle=0,width=0.9\columnwidth]{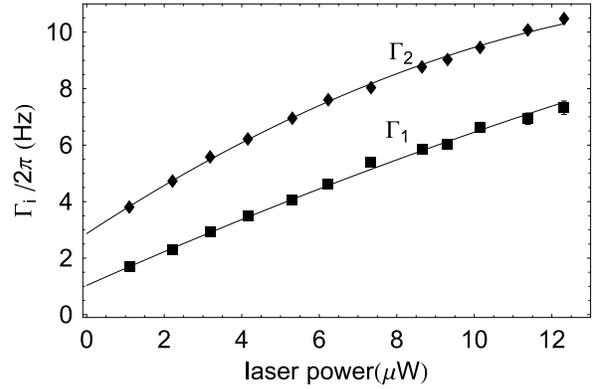}}
\caption{Laser power dependence of the relaxation rates $\sGamma{1}$
  (boxes) and $\sGamma{2}$ (diamonds).  The experimental points are
  fitted with (\protect\ref{eq:gammas}).  The (statistical) error bars
  on the individual data points are smaller than the symbol size.}
\label{fig:relaxation}
\end{figure}

\section{Results}
\label{sec:results}

\subsection{Relaxation rates}

Figure~\ref{fig:relaxation} shows the dependence of the longitudinal
and transverse relaxation rates on the laser power $P_L$.
There is, to our knowledge, no theoretical algebraic expression
describing that dependence for ground states of arbitrary angular
momentum $F$.
We therefore fit, as in \cite{DiDomenico:2007:SDR}, the dependence by
a quadratic polynomial
\begin{equation}
\sGamma{i}(P_L)= \sGamma{0i}+\alpha_{i}\,P_L+\beta_{i}\,P_L^2\,,
\label{eq:gammas}
\end{equation}
which allows us to infer the intrinsic relaxation rates, $\sGamma{01}$
and $\sGamma{02}$, i.e., the relaxation rates extrapolated to zero
light power.

\begin{figure}[b]
\centerline{\includegraphics[angle=0,width=0.9\columnwidth]{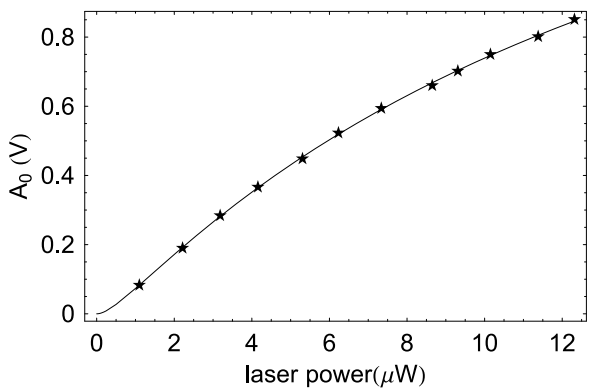}}
\caption{Magnetic resonance amplitude versus laser power.  The
  experimental points are fitted with the polynomial expression from
  (\protect\ref{eq:amlitude}), which yields, for this specific cell,
  the saturation parameters $P_{S1}=634$~nW and $P_{S2}=16.3~\mu$W.
  The error bars are smaller than the plotting symbol size.}
\label{fig:amplitude}
\end{figure}

\subsection{Signal amplitudes}

\begin{figure}[t!]
\centerline{\includegraphics[angle=0,width=0.85\columnwidth]{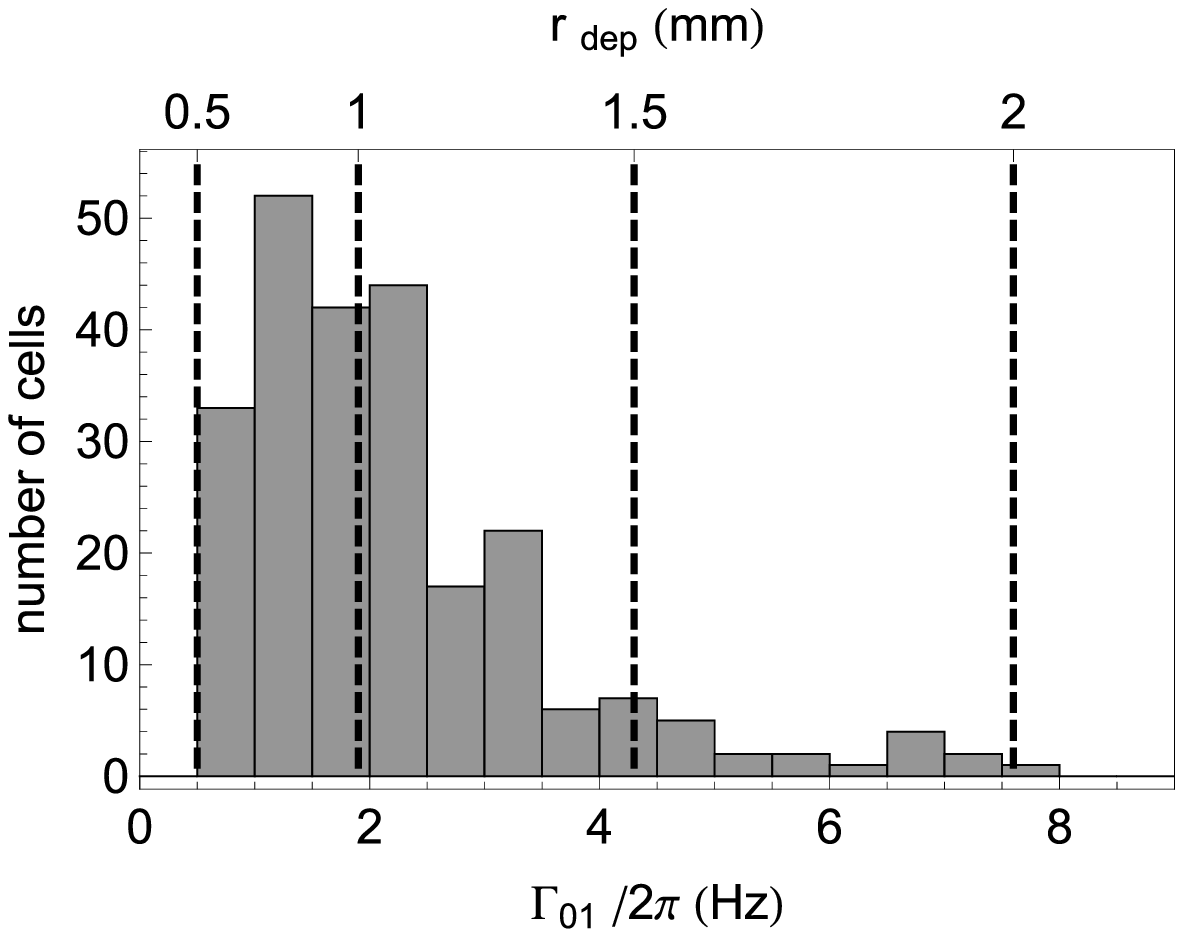}}
\centerline{\includegraphics[angle=0,width=0.85\columnwidth]{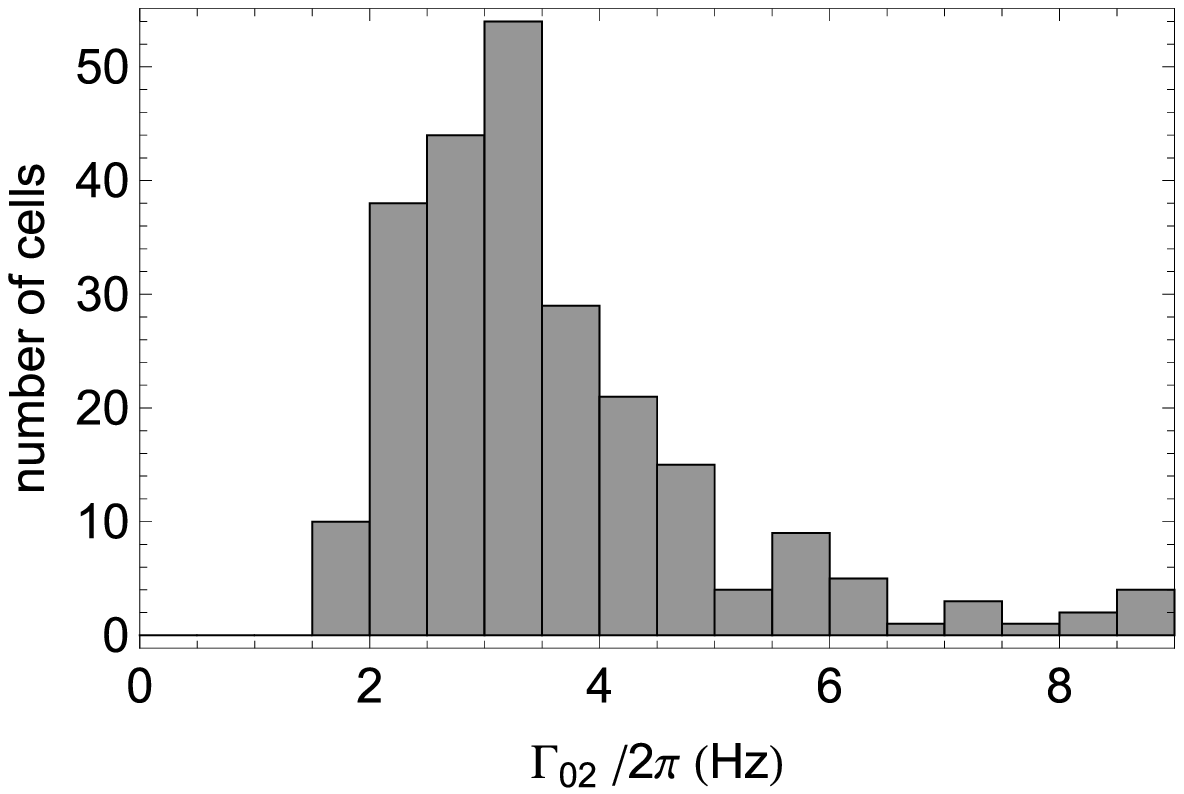}}
\centerline{\includegraphics[angle=0,width=0.85\columnwidth]{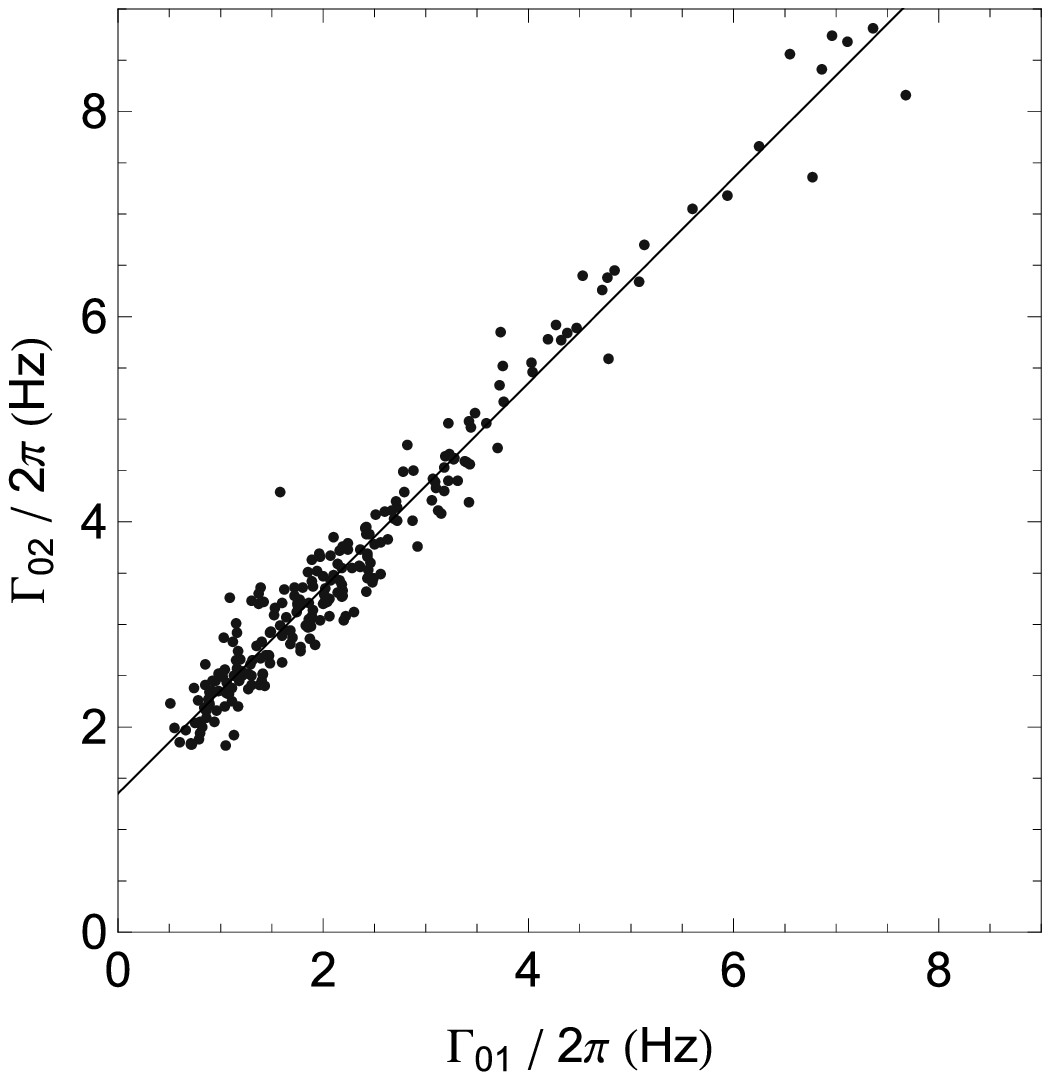}}
\caption{Histogram of intrinsic longitudinal (top) and transverse
  (middle) relaxation times of 241 coated cells.  The upper axis in
  the top graph gives the radius of the effective depolarization spot
  that models reservoir relaxation (see text).  The lower graph
  shows the correlation between the relaxation rates, together with a
  fit of the form $\sGamma{02} = s \sGamma{01} + a$.}
\label{fig:gamma12Histo}
\end{figure}

Figure~\ref{fig:amplitude} shows the dependence of the signal
amplitude $A_0$ on the laser power $P_L$.
Here again, we have no theoretically derived algebraic expression
describing that dependence for transitions between states with
arbitrary angular momenta.
We therefore, as in \cite{DiDomenico:2007:SDR}, fit the experimental
dependence by the empirical saturation formula
\begin{equation}
S_0(P_L)= C\,\frac{P_L^2}{(P_L+P_{S1})(P_L+P_{S2})}
\label{eq:amlitude}
\end{equation}
which accounts for an amplitude growing as $P_L^2$ at low powers, and
where $P_{S1}$ and $P_{S2}$ are saturation powers.

Figures.~\ref{fig:relaxation} and \ref{fig:amplitude} show typical
dependencies of $\sGamma{1}$, $\sGamma{2}$, and $S_0$ on $P_L$ for a
given cell, together with the fits (solid lines) by
(\ref{eq:amlitude}).
We have characterized 253 paraffin-coated cells of equal diameter
using the method described above.
The histograms in Fig.~\ref{fig:gamma12Histo} (top, middle) show the
distributions of the intrinsic longitudinal and transverse relaxation
rates of the 241 best cells.
The scatter plot in the lower graph of Fig.~\ref{fig:gamma12Histo}
shows that the two rates are strongly correlated.
The fitted line represents a linear relation of the form $\sGamma{02}=
s\,\sGamma{01} + a$ with $s=1.00(1)$ and $a=1.35(3)~\Hz{}$.
The longitudinal and transverse relaxation rates are thus equal, up to
a constant offset that affects the $\sGamma{02}$ values only.
For an isotropic relaxation process, in which all Zeeman sublevels
relax at the same rate, one would expect $\sGamma{01}=\sGamma{02}$.
In section~\ref{sec:discussion} below we will come back to a
quantitative discussion of those contributions.

\subsection{Magnetometric sensitivity}
\label{subsec:NEM}

The intrinsic relaxation rates are well suited to characterize each
individual cell.
In particular, the transverse rate $\sGamma{02}$, which determines the
intrinsic width of the signals $A_{\omega}$ and $D_{\omega}$, is
relevant for magnetometric applications.
However, the intrinsic rates are, by definition, rates for vanishing
laser and \rf{} powers.
Therefore, the magnetometric sensitivity of a given cell can not be
inferred directly from the intrinsic rates, since magnetometers have
to be operated at finite laser and \rf{} power levels.

The linear zero crossing of the dispersive signal $D_{\omega}$ near
resonance is convenient for magnetometric applications since any
magnetic field change $\delta B$ yields a signal change
\begin{equation}
\delta
D_{\omega}=\left|\frac{dD_{\omega}}{dB}|_{\omega{=}\omega_L}\right|\,
\delta B
\end{equation}
that is proportional to $\delta B$.
The lowest magnetic field change $\delta B$ that can be detected
depends on the shot noise of the DC photocurrent $I_{L}\propto P_L$.
A feedback resistor, $R_F$, in the transimpedance amplifier, marked
$I/U$ in Fig.~\ref{fig:expsetup}, transforms that photocurrent into a
photovoltage $U_L$, whose shot noise (in a bandwidth of 1~\Hz) is
given by
\begin{equation}
\delta U_{L} = R_F\,\delta I_{L}
             = R_F\,\sqrt{2 e I_{L}}
             = R_F\,\sqrt{\frac{2 \Qe P_L e^{2}}{h \nu}}\,,
\label{eq:Ushot}
\end{equation}
where $\Qe=70\%$ is the quantum efficiency of the
photodiode, and $\nu$ the laser frequency.
The experimentally measured signal noise lies $\approx 20\%$ above the
shot noise level, due to laser power fluctuations and amplifier noise.
With the calibration constant $\eta$ in (\ref{eq:signals}), $\delta
D_{\omega}$ is expressed in Volts, i.e., in the same units as $\delta
U_{L}$.

For each set of the experimental parameters $P_L$ and $\Omega_{\rf}$
one can thus define the magnetometric sensitivity as the field
fluctuation $\delta B_{\NEM}$ that induces a signal change $\delta
D_{\omega}$ of equal magnitude than $\delta U_{L}$.
This noise equivalent magnetic field fluctuation (\NEM) is thus given
by
\begin{eqnarray}
 \delta B_{\NEM}
 &=&\frac{\delta U_{L}} {\left|\frac{dD_{\omega}}{dB}|_{\omega{=}\omega_L}\right|}\\
 &=&\frac{1}{\gamma_F}
    \frac{\sGamma{2}^2+\Omega_{\rf}^2\sGamma{2}/\sGamma{1}}
         {A_0 \Omega_{rf}} \,\delta U_{L}\,.
 \label{eq:NEM}
\end{eqnarray}
$A_0$, $\sGamma{1}$, and $\sGamma{2}$ are ($P_L$ dependent) parameters
obtained from the fits of the experimental $D_\omega$ spectra. $\delta
U_L$ is assumed to be the $P_L$ dependent shot noise value
(\ref{eq:Ushot}).
We recall that $\Omega_{\rf}$ is not a fit parameter, and that
calibrated numerical values of $\Omega_{\rf}$ are inserted in
(\ref{eq:NEM}) when evaluating $\delta B_{\NEM}$.

\begin{figure}[t]
\centerline{\includegraphics[angle=0,width=0.9\linewidth]{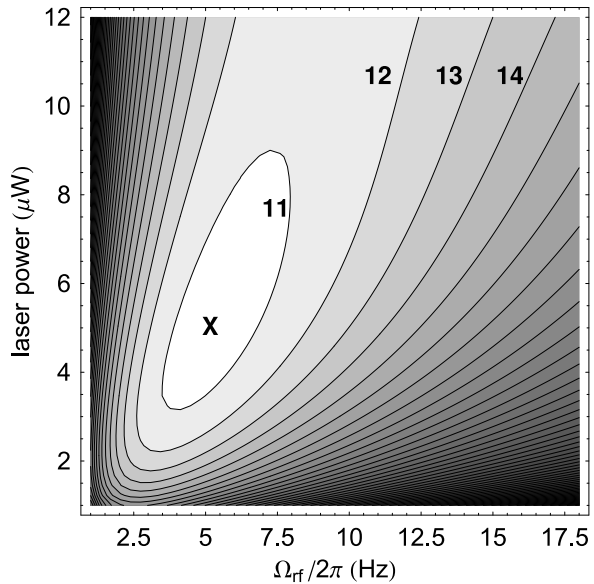}}
\caption{Plot of $\delta B_{\NEM}$ as a function of the amplitude
  $\Omega_{\rf}$ of the rotating field and of the laser power $P_L$.
  The contours represent the lines of constant \NEM, spaced by
  1~\fTHz, with selected numerical values indicated.  The cross refers
  to the minimal \NEM{} value, which, for the cell represented here
  has a value of 10.5~\fTHz.}
\label{fig:nem}
\end{figure}

For each cell we have evaluated $\delta B_{\NEM}$ for a range of
parameters $P_L$ and $\Omega_{\rf}$.
Figure~\ref{fig:nem} shows a typical result in terms of a contour plot
of $\delta B_{\NEM}$.
For each cell we determine the optimal \NEM{} value, $\delta
B_{\NEM}^\mathrm{min}$, by a numerical minimization procedure.
The minimum for the cell shown in Fig.~\ref{fig:nem} is indicated by a
cross.

\begin{figure}[t]
\centerline{\includegraphics[angle=0,width=0.95\linewidth]{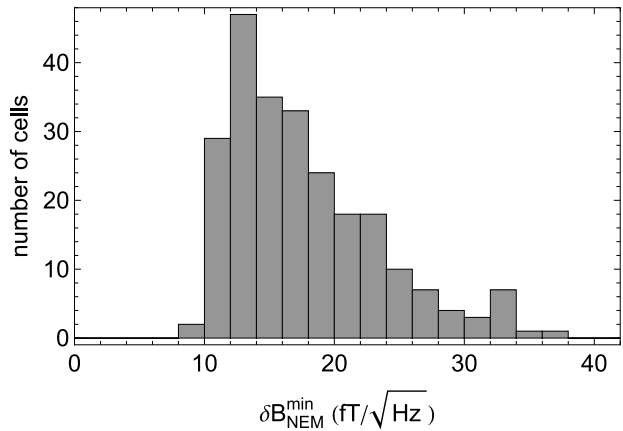}}
\caption{Histogram of the minimal \NEM{} values, $\delta
  B_{\NEM}^\mathrm{min}$, of 241 cells, which represent 94\% of
  the cells produced to date.}
\label{fig:histogram}
\end{figure}

The distribution of minimal \NEM{} values, $\delta
B_{\NEM}^\mathrm{min}$, thus obtained is represented in form of a
histogram in Fig.~\ref{fig:histogram}.
Only cells with $\delta B_{\NEM}^\mathrm{min}<40 \fTHz$ are
shown.
This set represents 94\% of all cells we have produced to date.

\section{Discussion}
\label{sec:discussion}

The distribution of linewidths shown in Fig.~\ref{fig:gamma12Histo}
reveals a dependence of the form $\sGamma{02} = \sGamma{01} + \Delta
\sGamma{\mathrm{relax}}$, with a constant offset relaxation rate $\Delta
\sGamma{\mathrm{relax}}$, whose numerical value (fit parameter $a$ in
Fig.~\ref{fig:gamma12Histo}) is $\Delta\sGamma{\mathrm{relax}} / 2\pi =
1.35~\Hz$.
Here we show that $\sGamma{01}$ is ultimately limited by atoms
escaping to the sidearm, and that $\Delta\sGamma{\mathrm{relax}}$ is
mainly determined by spin exchange collisions
($\Delta\sGamma{\mathrm{ex}}$) with a minor contribution from magnetic
field inhomogeneities ($\Delta\sGamma{\Delta B}$).

\subsection{Longitudinal relaxation}

The intrinsic longitudinal relaxation rate $\sGamma{01}$ is limited by
processes which thermalize the magnetic sublevel populations, such as
atoms escaping through the capillary to the sidearm where they
eventually collide with the solid Cs droplet, atoms hitting an
imperfectly coated surface spot of the spherical bulb, or atoms being
absorbed by the coating \cite{Alexandrov:2002:LID,Gozzini:2008:LIS}.
All of those processes can be parametrized in terms of an effective
depolarizing surface area $\sigma_\mathrm{dep} \equiv \pi
r_\mathrm{dep}^2$.
We will refer to such processes in general as ``reservoir losses''.
The distribution of $\sGamma{01}$ values in the top graph of
Fig.~\ref{fig:gamma12Histo} represents the statistical distribution of
such imperfections, due to uncontrolled parameters in the cell
production process.
In a spherical cell of radius $R$ the rate of wall collisions is
$\gamma_\mathrm{wall}=3\overline{v}/4R$, where $\overline{v}$ is the
average thermal velocity.
The intrinsic longitudinal relaxation rate can thus be expressed in
terms of the effective depolarizing spot radius, $r_\mathrm{dep}$, via
\begin{equation}
\sGamma{01} = \gamma_\mathrm{wall} \frac{\pi r_\mathrm{dep}^2}
                                        {4\pi R^2}
            = \frac{3\,\overline{v}\,r_\mathrm{dep}^2}
                   {16\,R^3}\,.
\end{equation}
The upper axis in the top graph of Fig.~\ref{fig:gamma12Histo} shows
the radius $r_\mathrm{dep}$ corresponding to the $\sGamma{01}$ value
on the lower axis.
The best cell produced so far has a longitudinal relaxation rate
$\sGamma{01}/ 2\pi\approx 0.50(5)~\Hz$, which corresponds to
$d_\mathrm{dep}=2r_\mathrm{dep}=1~\mm$.
This value is compatible with the design diameter, $d_\mathrm{cap} =
0.75(25)~\mm$, of the capillary, which shows that $\sGamma{01}$ is
ultimately limited by atoms escaping into the sidearm.

\subsection{Transverse relaxation: field inhomogeneities}

If the offset magnetic field $B_0$ varies over the cell volume it
produces a distribution of resonance frequencies $\omega_L$, and hence
a broadening of the magnetic resonance lines given by
(\ref{eq:signals}) and (\ref{eq:phase}).
The fitting analysis interprets this broadening as an increase of the
transverse linewidth $\sGamma{02}$ by an amount $\Delta\Gamma_{\Delta B}$.
A main advantage of coated cells over buffer gas filled cells is that,
because of multiple wall collisions, the atoms explore a large
fraction of the cell volume during the spin coherence time, which
effectively averages out field gradients.
Standard line narrowing theory \cite{Watanabe:1977:MLN} predicts that
an inhomogeneous magnetic field gives a lowest order contribution
\begin{equation}
\sGamma{\Delta B} = (\gamma_F \Delta B_\mathrm{rms})^2 \tau_c
\label{eq:gamma2inhom}
\end{equation}
to the transverse relaxation rate, where $\Delta B_\mathrm{rms}$ is
the rms value of the magnetic field averaged over the cell volume, and
$\tau_c$ the correlation time of the field fluctuations seen by the
cell, which can be approximated by the mean time between wall
collisions.
%
This expression is valid in the so-called good averaging regime
\cite{Watanabe:1977:MLN}, i.e., for $\gamma_F \Delta B_\mathrm{rms}
\tau_c \ll 1$.
From the geometry of the used coils we estimate $\Delta
B_\mathrm{rms}$ to be on the order of 2~\nT, which yields
$\Delta\nu_{\Delta B}=\Gamma_{\Delta B}/2\pi= 30$~mHz.
Even when allowing for a 5~times larger inhomogeneity (i.e., $\Delta
B=10~\nT$) from uncompensated residual fields --- recall that we
actively compensate linear field gradients --- one still has
$\Delta\nu_{\Delta B}< 0.1~\Hz$.  We can thus ascertain that the
contribution from field inhomogeneities to $\sGamma{02}$ is
negligible.
We note that the good averaging conditions for $\Delta B=2~\nT$ and
10~\nT{} read $\gamma_F\Delta B_\mathrm{rms}\tau_c=0.004$ and $0.02$,
respectively.

\subsection{Transverse relaxation: spin exchange}

As derived by Ressler~\etal, \cite{Ressler:1969:MSE}, the
contribution from spin exchange collisions to the transverse
relaxation rate is given by
\begin{equation}
\Delta\sGamma{\mathrm{ex}}=\alpha\frac{2I}{2I+1}\,
                           n_\mathrm{Cs}\sigma_\mathrm{ex}v_r\,,
\label{eq:sex}
\end{equation}
where $I$ is the nuclear spin, $n_\mathrm{Cs}$ the Cs number density,
$v_r$ the relative velocity of colliding atoms, and
$\sigma_\mathrm{ex} = 2.06\times 10 ^{-14}~\mathrm{cm}^2$
\cite{Ressler:1969:MSE} the spin exchange cross section for Cs--Cs
collisions.
The parameter $\alpha$ describes the slowing down of the spin
relaxation due to the hyperfine interaction.
In small magnetic fields $\alpha\approx 0.63$ for the
$M{=}-4\rightarrow M{=}-3$ transition (Fig.~3 of
\cite{Ressler:1969:MSE}).
At $T=20(1){}\,^\circ$C the contribution of spin exchange collisions
to $\sGamma{02}$ evaluates to
\begin{equation}
\frac{\Delta\sGamma{\mathrm{ex}}}{2\pi}= 1.6(2)~\Hz \,,
\label{eq:sexval}
\end{equation}
where the error reflects the uncertainty in the number density.
This value is compatible with the experimental value
$\sGamma{02}-\sGamma{01} = (2\pi) 1.35(3) ~\Hz$.
We are therefore confident that dephasing spin exchange collisions
give the main contribution to the transverse relaxation rate,
notwithstanding a certain scatter of the spin exchange cross sections
in the literature.

\subsection{Fundamental limits of magnetometric sensitivity}

The ultimate sensitivity of the type of magnetometers described here
is limited by two fundamental processes, viz., photon shot noise limit
and spin projection noise.
One can show (Appendix~\ref{sec:photoncounting}) that the minimal
\NEM{} imposed by the shot noise of the detected photons is given by
\begin{equation}
\delta B_{\NEM}^\mathrm{PSN}
=
\frac{2\sqrt{2}\sGamma{2}}{\gamma_F}
\sqrt{\frac{\sGamma{2}}{\sGamma{1}}}
\frac{1}{\kappa_0 L }
\frac{1}{\AnaFF\Fpol}
\sqrt{\frac{h \nu}{\Qe \Pout t}}
\,,
\label{eq:PSN}
\end{equation}
where \Pout{} is the power detected after the cell, \Qe{} the quantum
efficiency of the photodiode for photons of energy $h\nu$, and
$\kappa_0$ the resonant absorption coefficient of the driven hyperfine
component for unpolarized atoms.
For a time interval of $t=0.5$~s, the result corresponds to a
measurement bandwidth of 1~\Hz.
In (\ref{eq:PSN}) \Fpol{} is the spin polarization
\begin{equation}
\Fpol = \sum_{M=-4}^{4} p_{4,M}^{\phantom{+}} M\,,
\label{eq:Fz}
\end{equation}
in the $F{=}4$ state, where the $p_{4,M}$ are the populations of the
magnetic sublevels $\ket{F{=}4,M}$.
The analyzing power for the transition $F{\rightarrow}F'$, \AnaFF{},
depends in general on the applied laser power and accounts for
population effects such as hyperfine pumping.  Its value has been
determined by a numerical model based on rate equations
\cite{pumping_inprep}.  It is a slowly varying function in the domain
of laser powers considered here, with value $\AnaFFus = 1.15(5)$.

For our apparatus, $\kappa_0 L\approx 0.7$, $\Qe=0.7$, so the above
can be rewritten as
\begin{equation}
\delta B_{\NEM}^\mathrm{PSN}(\fT) = \frac{0.146}{\AnaFFus\Fpol\sqrt{\Pout(\mu\mbox{W})}}
                                    \sGamma{2}\sqrt{\frac{\sGamma{2}}{\sGamma{1}}} \,.
\label{eq:PSNnumerical}
\end{equation}
For our best cell, $\AnaFFus\Fpol = 0.39(4)$, $\sGamma{1}/2\pi =
3.40~\Hz$, and $\sGamma{2}/2\pi = 4.75~\Hz$ at the optimum laser power
of $3.6~\mu$W, which yields an expected sensitivity of $\delta
B_{\NEM}^\mathrm{PSN}=7.0(7)~\fT$, to be compared with the measured
minimal \NEM{} of the cell of 9(1)~\fT.  For a more typical cell with
$\sGamma{2}/2\pi = 10~\Hz$, $\sGamma{1}/2\pi = 8.65~\Hz$, and
$\AnaFFus\Fpol = 0.46(5)$ at the optimal power of $5~\mu$W, the
expected minimal \NEM{} is $\delta B_{\NEM}^\mathrm{PSN} =
9.6(1.0)~\fT$, indicating that the shot noise limited \NEM{} grows
less than linearly in $\sGamma{2}$.

Spin projection noise limits the magnetometric sensitivity to
\begin{equation}
\delta B_{\NEM}^\mathrm{SPN}=
\frac{1}{\gamma_F} \sqrt{\frac{\sGamma{2}}
                              {N_\mathrm{at} t_\mathrm{meas}}}\,,
\label{eq:SPN}
\end{equation}
where $N_\mathrm{at}=\frac{9}{16}\rho_\mathrm{at} V_\mathrm{cell}$ is
the number of atoms in the $F{=}4$ state that contribute to the signal,
with $\rho_\mathrm{at}$ being the total Cs number density, and
$V_\mathrm{cell}$ the cell volume.
For a measurement time $t_\mathrm{meas}$ of 0.5~s, one finds at
$T=20(1){}\,^\circ$C, $\delta B_{\NEM}^\mathrm{SPN}=0.74(2)~\fT$ for
$\sGamma{2}/2\pi= 4.75~\Hz$.  In our magnetometers spin projection noise
thus has a negligible contribution.

\section{Summary and conclusion}
\label{sec:conclusion}

We have manufactured and characterized a set of 253 paraffin-coated Cs
vapor cells of identical geometry (15~mm radius spheres), 90\% of
which have an intrinsic transverse relaxation rate in the range of
2~to 6~\Hz.
Under optimized conditions of laser and \rf{} power those cells have
intrinsic magnetometric sensitivities, $\delta B_{\NEM}^\mathrm{min}$,
in the range of 9 to 30~\fTHz{} under the assumption of (light)
shot-noise limited operation in a DROM-type magnetometer.

The magnetometric sensitivity is determined by the intrinsic
transverse relaxation rate, which, for the best cell of our batch has
a value of $2\pi\cdot2~\Hz$, of which $\approx~0.5~\Hz$ are due to
reservoir ($T_1$) relaxation, and $\approx~1.5~\Hz$ are due to spin
exchange relaxation.
Improving the relaxation properties by reducing reservoir relaxation
is technologically demanding, and would only marginally improve the
overall sensitivity.
Spin exchange relaxation, on the other hand, cannot be suppressed in
coated cells, although it was shown that spin exchange relaxation can
be suppressed in high pressure buffer gas cells, yielding sub-\fT{}
magnetometric sensitivity \cite{Kominis:2003:SMA}.
We thus conclude that our cells are as good as coated cells of that
diameter can be, disregarding a possible 25\% reduction of
$\sGamma{02}/2\pi$ by a suppression of reservoir losses.

The expected photon shot noise limited \NEM{} of our cells is very
close to the measured \NEM{}.  The most promising improvement in
sensitivity is expected to come from maximizing \Fpol{} via hyperfine
repumping, which could win, at most, a factor of 2--3.

It is well known that in the spin exchange limited regime an increase
of the atomic density by heating the cell does not increase the
magnetometric sensitivity, since both $\sGamma{2}$ and $\kappa_0$ in
(\ref{eq:PSN}) grow proportionally to the density.
The same holds for $\sGamma{2}$ and $N_\mathrm{atom}$ in
(\ref{eq:SPN}).
However, when operating the magnetometer in a regime where spin
exchange is not the limiting factor, one expects an improvement of the
sensitivity by increasing the atomic number density.

We will use the cells in multi-sensor applications in fundamental and
applied fields of research.
Since an optimal magnetometric sensitivity is reached with a typical
light power of approximately $5~\mu$W, a single diode laser can drive
hundreds of individual sensors \cite{Hofer:2008:HSO}.
This scalability, together with the very good reproducibility of the coated
cell quality reported here, will allow us to realize in the near future
a three-dimensional array of 25 individual sensors for imaging the
magnetic field of the beating human heart, a signal with a peak amplitude
100~\pT{} \cite{Hofer:2008:HSO}.
With a reliable and inexpensive multichannel heart measurement system,
magnetocardiograms can be measured in a few minutes, times which are
of interest in the real world of clinical applications.

\appendix
\section{Photon shot noise limit}
\label{sec:photoncounting}

Consider a light beam of power \Pin{} traversing a vapor of thickness
$L$.  The transmitted power detected by a photodiode with quantum
efficiency \Qe{} is given by
\begin{equation}
\Pdet(t) =\Qe\Pin e^{-\kappa(t) L}=\Qe\Pout\,.
\label{eq:beer}
\end{equation}
The time dependent absorption coefficient $\kappa(t)$ considering the
in-phase component of the magnetic-resonance induced modulation is
\begin{equation}
\kappa(t) = \kappa_0 \left( 1  - \AnaFF \Fpol
   \frac {\Omega_{\rf} \delta }
   { \delta^2+ \sGamma{2}^{2}+\frac{\sGamma{2}}{\sGamma{1}} \Omega^2_{\rf} }
\cos{\omega t}
 \right)\,,
\label{eq:kapparest}
\end{equation}
where $\kappa_0$ is the resonant optical absorption coefficient for a
sample of unpolarized atoms and the polarization \Fpol{} is as defined
by (\ref{eq:Fz}).
The analyzing power for transition $F{\rightarrow}F'$, \AnaFF{},
accounts for population effects such as hyperfine pumping: more
details are given in the discussion in the main text following
(\ref{eq:Fz}).
Lock-in detection extracts from (\ref{eq:beer}) the rms value
\begin{equation}
\PAC = \frac{1}{\sqrt{2}} \Qe \Pin e^{-\kappa_0 L}
\: \kappa_0 L \AnaFF\Fpol \frac {\Omega_{\rf} \delta }
   {\delta^2+ \sGamma{2}^{2}+\frac{\sGamma{2}}{\sGamma{1}}
     \Omega^2_{\rf}}
\label{eq:kappares}
\end{equation}
of the in-phase component of the power modulation.

Light power $P$ can be converted to a photon count $N$ seen during
time $t$ via $N = \frac{P t}{h \nu}$, for $\nu$ the photon frequency.
Hence \NAC{} represents the number of photons carrying magnetometric
information, thus
\begin{equation}
\frac{d\NAC}{dB} = \frac{d\NAC}{d\PAC}
\frac{d\PAC}{d\delta}
\frac{d\delta}{dB}\,,
\label{eq:chainrule}
\end{equation}
which evaluates to
\begin{equation}
\frac{d\NAC}{dB} =
\frac{1}{\sqrt{2}}
\: \kappa_0 L e^{-\kappa_0 L} \:
\Nin
\Qe  \AnaFF\Fpol \frac{\gamma_F}{2\sGamma{2}}
\sqrt{\frac{\sGamma{1}}{\sGamma{2}}}
\,,
\label{eq:thederiv}
\end{equation}
assuming an \rf{} amplitude ($\Omega_{\rf} =
\sqrt{\sGamma{2}\sGamma{1}}$) which maximizes the result.

The photon shot noise limited magnetometric sensitivity is given by
\begin{equation}
\delta B_{\NEM}^\mathrm{PSN} =
\sqrt{\NDC} \left( \frac{d\NAC}{dB} \right)^{-1}\,,
\label{eq:limit}
\end{equation}
where
\begin{equation}
\NDC = \left< \Qe \Nin e^{-\kappa(t) L} \right >_{t} = \Qe \Nin e^{-\kappa_0 L}\,.
\label{eq:ndcdef}
\end{equation}
Assembling the above components gives
\begin{equation}
\delta B_{\NEM}^\mathrm{PSN}  =
\frac{2\sqrt{2}\sGamma{2}}{\gamma_F}
\sqrt{\frac{\sGamma{2}}{\sGamma{1}}}
\frac{1}{\kappa_0 L }
\frac{1}{\AnaFF\Fpol}
\sqrt{\frac{h \nu}{\Qe \Pout t}}
\,,
\end{equation}
which is the required result.

\vspace*{2cm}

This work is financially supported by the Swiss National Science
Foundation (\#200020--111958, \#200020-119820, \#200020--113641) and
by the Velux Foundation.


\begin{raggedright}

\begin{thebibliography}{} 
\expandafter\ifx\csname natexlab\endcsname\relax\def\natexlab#1{#1}\fi
\expandafter\ifx\csname bibnamefont\endcsname\relax
  \def\bibnamefont#1{#1}\fi
\expandafter\ifx\csname bibfnamefont\endcsname\relax
  \def\bibfnamefont#1{#1}\fi
\expandafter\ifx\csname citenamefont\endcsname\relax
  \def\citenamefont#1{#1}\fi
\expandafter\ifx\csname url\endcsname\relax
  \def\url#1{\texttt{#1}}\fi
\expandafter\ifx\csname urlprefix\endcsname\relax\def\urlprefix{URL }\fi
\providecommand{\bibinfo}[2]{#2}
\providecommand{\eprint}[2][]{\url{#2}}

\bibitem{Budker:2002:RNM}
\bibinfo{author}{\bibfnamefont{D.} \bibnamefont{Budker}},
  \bibinfo{author}{\bibfnamefont{W.} \bibnamefont{Gawlik}},
  \bibinfo{author}{\bibfnamefont{D.~F.} \bibnamefont{Kimball}},
  \bibinfo{author}{\bibfnamefont{S.~M.} \bibnamefont{Rochester}},
  \bibnamefont{and} \bibinfo{author}{\bibfnamefont{V.~V.}
  \bibnamefont{Yashchuk}},
  \bibinfo{author}{\bibfnamefont{A.} \bibnamefont{Weis}},
  \bibinfo{journal}{Reviews of Modern Physics}
  \textbf{\bibinfo{volume}{74}}, \bibinfo{pages}{1153} (\bibinfo{year}{2002}).

\bibitem{Robinson:1958:PSS}
\bibinfo{author}{\bibfnamefont{H.~G.} \bibnamefont{Robinson}},
  \bibinfo{author}{\bibfnamefont{E.~S.} \bibnamefont{Ensberg}},
  \bibnamefont{and} \bibinfo{author}{\bibfnamefont{H.~G.}
  \bibnamefont{Dehmelt}}, \bibinfo{journal}{Bull. Am. Phys. Soc.}
  \textbf{\bibinfo{volume}{3}} (\bibinfo{year}{1958}).

\bibitem{DiDomenico:2007:SDR}
\bibinfo{author}{\bibfnamefont{G.}~\bibnamefont{{Di~Domenico}}},
  \bibinfo{author}{\bibfnamefont{H.}~\bibnamefont{Saudan}},
  \bibinfo{author}{\bibfnamefont{G.}~\bibnamefont{Bison}},
  \bibinfo{author}{\bibfnamefont{P.}~\bibnamefont{Knowles}}, \bibnamefont{and}
  \bibinfo{author}{\bibfnamefont{A.}~\bibnamefont{Weis}},
  \bibinfo{journal}{Phys.~Rev.~A} \textbf{\bibinfo{volume}{76}},
  \bibinfo{eid}{023407} (pages~\bibinfo{numpages}{10}) (\bibinfo{year}{2007}),
  \urlprefix\url{http://link.aps.org/abstract/PRA/v76/e023407}.

\bibitem{Weis:2005:LBP}
\bibinfo{author}{\bibfnamefont{A.}~\bibnamefont{Weis}} \bibnamefont{and}
  \bibinfo{author}{\bibfnamefont{R.}~\bibnamefont{Wynands}},
  \bibinfo{journal}{Opt.~Las.~Engineer.} \textbf{\bibinfo{volume}{43}},
  \bibinfo{pages}{387} (\bibinfo{year}{2005}).

\bibitem{Budker:2007:OM}
\bibinfo{author}{\bibfnamefont{D.}~\bibnamefont{Budker}} \bibnamefont{and}
  \bibinfo{author}{\bibfnamefont{M.}~\bibnamefont{Romalis}},
  \bibinfo{journal}{Nature Physics} \textbf{\bibinfo{volume}{3}},
  \bibinfo{pages}{227} (\bibinfo{year}{2007}),
  \urlprefix\url{http://dx.doi.org/10.1038/nphys566}.

\bibitem{Klein:2006:SLP}
\bibinfo{author}{\bibfnamefont{M.} \bibnamefont{Klein}},
  \bibinfo{author}{\bibfnamefont{I.} \bibnamefont{Novikova}},
  \bibinfo{author}{\bibfnamefont{D.~F.} \bibnamefont{Phillips}},
  \bibnamefont{and} \bibinfo{author}{\bibfnamefont{R.~L.}
  \bibnamefont{Walsworth}},
  \bibinfo{journal}{Journal of Modern Optics}
  \textbf{\bibinfo{volume}{53}}, \bibinfo{pages}{2583} (\bibinfo{year}{2006}).

\bibitem{Fernholz:2008:SSA}
\bibinfo{author}{\bibfnamefont{T.}~\bibnamefont{Fernholz}},
  \bibinfo{author}{\bibfnamefont{H.}~\bibnamefont{Krauter}},
  \bibinfo{author}{\bibfnamefont{K.}~\bibnamefont{Jensen}},
  \bibinfo{author}{\bibfnamefont{J.~F.} \bibnamefont{Sherson}},
  \bibinfo{author}{\bibfnamefont{A.~S.} \bibnamefont{Sorensen}},
  \bibnamefont{and} \bibinfo{author}{\bibfnamefont{E.~S.}
  \bibnamefont{Polzik}}, \bibinfo{journal}{Physical Review Letters}
  \textbf{\bibinfo{volume}{101}}, \bibinfo{eid}{073601}
  (pages~\bibinfo{numpages}{4}) (\bibinfo{year}{2008}),
  \urlprefix\url{http://link.aps.org/abstract/PRL/v101/e073601}.

\bibitem{Alexandrov:2002:LID}
\bibinfo{author}{\bibfnamefont{E.~B.} \bibnamefont{Alexandrov}},
  \bibinfo{author}{\bibfnamefont{M.~V.} \bibnamefont{Balabas}},
  \bibinfo{author}{\bibfnamefont{D.}~\bibnamefont{Budker}},
  \bibinfo{author}{\bibfnamefont{D.}~\bibnamefont{English}},
  \bibinfo{author}{\bibfnamefont{D.~F.} \bibnamefont{Kimball}},
  \bibinfo{author}{\bibfnamefont{C.-H.} \bibnamefont{Li}}, \bibnamefont{and}
  \bibinfo{author}{\bibfnamefont{V.~V.} \bibnamefont{Yashchuk}},
  \bibinfo{journal}{Phys.~Rev.~A} \textbf{\bibinfo{volume}{66}},
  \bibinfo{pages}{042903} (\bibinfo{year}{2002}),
  \urlprefix\url{http://prola.aps.org/abstract/PRA/v66/i4/e042903}.

\bibitem{Gozzini:2008:LIS}
\bibinfo{author}{\bibfnamefont{S.}~\bibnamefont{Gozzini}},
  \bibinfo{author}{\bibfnamefont{A.}~\bibnamefont{Lucchesini}},
  \bibinfo{author}{\bibfnamefont{L.}~\bibnamefont{Marmugi}}, \bibnamefont{and}
  \bibinfo{author}{\bibfnamefont{G.}~\bibnamefont{Postorino}},
  \bibinfo{journal}{The European Physical Journal D}
  \textbf{\bibinfo{volume}{47}}, \bibinfo{pages}{1} (\bibinfo{year}{2008}),
  \urlprefix\url{http://dx.doi.org/doi/10.1140/epjd/e2008-00015-5}.

\bibitem{Bison:2005:OPO}
\bibinfo{author}{\bibfnamefont{G.}~\bibnamefont{Bison}},
  \bibinfo{author}{\bibfnamefont{R.}~\bibnamefont{Wynands}}, \bibnamefont{and}
  \bibinfo{author}{\bibfnamefont{A.}~\bibnamefont{Weis}},
  \bibinfo{journal}{J.~Opt.~Soc.~Am.~B.} \textbf{\bibinfo{volume}{22}},
  \bibinfo{pages}{77} (\bibinfo{year}{2005}).

\bibitem{Bison:2003:LPM}
\bibinfo{author}{\bibfnamefont{G.}~\bibnamefont{Bison}},
  \bibinfo{author}{\bibfnamefont{R.}~\bibnamefont{Wynands}}, \bibnamefont{and}
  \bibinfo{author}{\bibfnamefont{A.}~\bibnamefont{Weis}},
  \bibinfo{journal}{Appl.~Phys.~B} \textbf{\bibinfo{volume}{76}},
  \bibinfo{pages}{325} (\bibinfo{year}{2003}).

\bibitem{Weis:2006:TDR}
\bibinfo{author}{\bibfnamefont{A.}~\bibnamefont{Weis}},
  \bibinfo{author}{\bibfnamefont{G.}~\bibnamefont{Bison}}, \bibnamefont{and}
  \bibinfo{author}{\bibfnamefont{A.~S.} \bibnamefont{Pazgalev}},
  \bibinfo{journal}{Phys.~Rev.~A} \textbf{\bibinfo{volume}{74}},
  \bibinfo{eid}{033401} (pages~\bibinfo{numpages}{8}) (\bibinfo{year}{2006}),
  \urlprefix\url{http://link.aps.org/abstract/PRA/v74/e033401}.

\bibitem{Hofer:2008:HSO}
\bibinfo{author}{\bibfnamefont{A.}~\bibnamefont{Hofer}},
  \bibinfo{author}{\bibfnamefont{G.}~\bibnamefont{Bison}},
  \bibinfo{author}{\bibfnamefont{N.}~\bibnamefont{Castagna}},
  \bibinfo{author}{\bibfnamefont{P.}~\bibnamefont{Knowles}},
  \bibinfo{author}{\bibfnamefont{J.~L.} \bibnamefont{Schenker}},
  \bibnamefont{and} \bibinfo{author}{\bibfnamefont{A.}~\bibnamefont{Weis}},
  \bibinfo{journal}{In Preparation}  (\bibinfo{year}{2008}).

\bibitem{Groeger:2006:HSL}
\bibinfo{author}{\bibfnamefont{S.}~\bibnamefont{Groeger}},
  \bibinfo{author}{\bibfnamefont{G.}~\bibnamefont{Bison}},
  \bibinfo{author}{\bibfnamefont{J.-L.} \bibnamefont{Schenker}},
  \bibinfo{author}{\bibfnamefont{R.}~\bibnamefont{Wynands}}, \bibnamefont{and}
  \bibinfo{author}{\bibfnamefont{A.}~\bibnamefont{Weis}},
  \bibinfo{journal}{Eur.~Phys.~J.~D} \textbf{\bibinfo{volume}{38}},
  \bibinfo{pages}{239} (\bibinfo{year}{2006}).

\bibitem{Ban:2006:TNM}
\bibinfo{author}{\bibfnamefont{G.}~\bibnamefont{{Ban}}},
  \bibinfo{author}{\bibfnamefont{K.}~\bibnamefont{{Bodek}}},
  \bibinfo{author}{\bibfnamefont{M.}~\bibnamefont{{Daum}}},
  \bibinfo{author}{\bibfnamefont{R.}~\bibnamefont{{Henneck}}},
  \bibinfo{author}{\bibfnamefont{S.}~\bibnamefont{{Heule}}},
  \bibinfo{author}{\bibfnamefont{M.}~\bibnamefont{{Kasprzak}}},
  \bibinfo{author}{\bibfnamefont{N.}~\bibnamefont{{Khomytov}}},
  \bibinfo{author}{\bibfnamefont{K.}~\bibnamefont{{Kirch}}},
  \bibinfo{author}{\bibfnamefont{A.}~\bibnamefont{{Knecht}}},
  \bibinfo{author}{\bibfnamefont{S.}~\bibnamefont{{Kistryn}}},
  \bibnamefont{et~al.}, \bibinfo{journal}{Hyperfine Interactions}
  \textbf{\bibinfo{volume}{172}}, \bibinfo{pages}{41} (\bibinfo{year}{2006}).

\bibitem{Andalkar:2002:HRM}
\bibinfo{author}{\bibfnamefont{A.}~\bibnamefont{Andalkar}} \bibnamefont{and}
  \bibinfo{author}{\bibfnamefont{R.~B.} \bibnamefont{Warrington}},
  \bibinfo{journal}{Phys.~Rev.~A} \textbf{\bibinfo{volume}{65}},
  \bibinfo{pages}{032708} (\bibinfo{year}{2002}).

\bibitem{Vanier:1989:QPA}
\bibinfo{author}{\bibfnamefont{J.}~\bibnamefont{Vanier}} \bibnamefont{and}
  \bibinfo{author}{\bibfnamefont{C.}~\bibnamefont{Audoin}},
  \emph{\bibinfo{title}{The Quantum Physics of Atomic Frequency Standards}}
  (\bibinfo{publisher}{Adam Hilger, Bristol and Philadelphia},
  \bibinfo{year}{1989}).

\bibitem{Budker:2005:MTN}
\bibinfo{author}{\bibfnamefont{D.}~\bibnamefont{Budker}},
  \bibinfo{author}{\bibfnamefont{L.}~\bibnamefont{Hollberg}},
  \bibinfo{author}{\bibfnamefont{D.~F.} \bibnamefont{Kimball}},
  \bibinfo{author}{\bibfnamefont{J.}~\bibnamefont{Kitching}},
  \bibinfo{author}{\bibfnamefont{S.}~\bibnamefont{Pustelny}}, \bibnamefont{and}
  \bibinfo{author}{\bibfnamefont{V.~V.} \bibnamefont{Yashchuk}},
  \bibinfo{journal}{Phys.~Rev.~A} \textbf{\bibinfo{volume}{71}},
  \bibinfo{eid}{012903} (pages~\bibinfo{numpages}{9}) (\bibinfo{year}{2005}),
  \urlprefix\url{http://link.aps.org/abstract/PRA/v71/e012903}.

\bibitem{Aleksandrov:1995:LPS}
\bibinfo{author}{\bibfnamefont{E.~B.} \bibnamefont{Aleksandrov}},
  \bibinfo{author}{\bibfnamefont{M.~V.} \bibnamefont{Balabas}},
  \bibinfo{author}{\bibfnamefont{A.~K.} \bibnamefont{Vershovskii}},
  \bibinfo{author}{\bibfnamefont{A.~E.} \bibnamefont{Ivanov}},
  \bibinfo{author}{\bibfnamefont{N.~N.} \bibnamefont{Yakobson}},
  \bibinfo{author}{\bibfnamefont{V.~L.} \bibnamefont{Velichanskii}},
  \bibnamefont{and} \bibinfo{author}{\bibfnamefont{N.~V.}
  \bibnamefont{Senkov}}, \bibinfo{journal}{Opt.~Spectrosc.}
  \textbf{\bibinfo{volume}{78}}, \bibinfo{pages}{325} (\bibinfo{year}{1995}).

\bibitem{Corwin:1998:FSD}
\bibinfo{author}{\bibfnamefont{K.~L.} \bibnamefont{Corwin}},
  \bibinfo{author}{\bibfnamefont{Z.~T.} \bibnamefont{Lu}},
  \bibinfo{author}{\bibfnamefont{C.~F.} \bibnamefont{Hand}},
  \bibinfo{author}{\bibfnamefont{R.~J.} \bibnamefont{Epstain}},
  \bibnamefont{and} \bibinfo{author}{\bibfnamefont{C.~E.}
  \bibnamefont{Wieman}}, \bibinfo{journal}{Appl.~Opt.}
  \textbf{\bibinfo{volume}{37}}, \bibinfo{pages}{3295} (\bibinfo{year}{1998}).

\bibitem{DiDomenico:2006:ESL}
\bibinfo{author}{\bibfnamefont{G.}~\bibnamefont{{Di~Domenico}}},
  \bibinfo{author}{\bibfnamefont{G.}~\bibnamefont{Bison}},
  \bibinfo{author}{\bibfnamefont{S.}~\bibnamefont{Groeger}},
  \bibinfo{author}{\bibfnamefont{P.}~\bibnamefont{Knowles}},
  \bibinfo{author}{\bibfnamefont{A.~S.} \bibnamefont{Pazgalev}},
  \bibinfo{author}{\bibfnamefont{M.}~\bibnamefont{Rebetez}},
  \bibinfo{author}{\bibfnamefont{H.}~\bibnamefont{Saudan}}, \bibnamefont{and}
  \bibinfo{author}{\bibfnamefont{A.}~\bibnamefont{Weis}},
  \bibinfo{journal}{Phys.~Rev.~A} \textbf{\bibinfo{volume}{74}},
  \bibinfo{eid}{063415} (pages~\bibinfo{numpages}{8}) (\bibinfo{year}{2006}),
  \urlprefix\url{http://link.aps.org/abstract/PRA/v74/e063415}.

\bibitem{Mathematica52}
\bibinfo{author}{\bibnamefont{{Wolfram Research, Inc.}}},
  \emph{\bibinfo{title}{Mathematica, {V5.2}}} (\bibinfo{publisher}{Wolfram
  Research, Inc., Champaign, Illinois}, \bibinfo{year}{2008}).

\bibitem{Watanabe:1977:MLN}
\bibinfo{author}{\bibfnamefont{S.~F.} \bibnamefont{Watanabe}} \bibnamefont{and}
  \bibinfo{author}{\bibfnamefont{H.~G.} \bibnamefont{Robinson}},
  \bibinfo{journal}{J.~Phys.~B--At.~Mol.~Opt.~Phys.}
  \textbf{\bibinfo{volume}{10}}, \bibinfo{pages}{931} (\bibinfo{year}{1977}).

\bibitem{Ressler:1969:MSE}
\bibinfo{author}{\bibfnamefont{N.~W.} \bibnamefont{Ressler}},
  \bibinfo{author}{\bibfnamefont{R.~H.} \bibnamefont{Sands}}, \bibnamefont{and}
  \bibinfo{author}{\bibfnamefont{T.~E.} \bibnamefont{Stark}},
  \bibinfo{journal}{Phys.~Rev.} \textbf{\bibinfo{volume}{184}},
  \bibinfo{pages}{102} (\bibinfo{year}{1969}).

\bibitem{pumping_inprep}
\bibinfo{author}{\bibfnamefont{A.}~\bibnamefont{Weis}} \bibnamefont{et~al.}
  (\bibinfo{year}{2009}), \bibinfo{note}{article in preparation}.

\bibitem{Kominis:2003:SMA}
\bibinfo{author}{\bibfnamefont{I.~K.} \bibnamefont{Kominis}},
  \bibinfo{author}{\bibfnamefont{T.~W.} \bibnamefont{Kornack}},
  \bibinfo{author}{\bibfnamefont{J.~C.} \bibnamefont{Allred}},
  \bibnamefont{and} \bibinfo{author}{\bibfnamefont{M.~V.}
  \bibnamefont{Romalis}}, \bibinfo{journal}{Nature}
  \textbf{\bibinfo{volume}{422}}, \bibinfo{pages}{596} (\bibinfo{year}{2003}).





\end{thebibliography}

\end{raggedright}

\end{document}